\documentclass[reprint,prl,a4paper,twocolumn,superscriptaddress,showpacs]{revtex4-1}
\usepackage{amsfonts}
\usepackage{amsmath}
\usepackage{amssymb}
\usepackage{graphicx}
\usepackage{hyperref}
\usepackage{dcolumn}

\begin{document}

\title{Experimentally witnessing the quantumness of correlations}

\author{R. Auccaise}
\affiliation{Empresa Brasileira de Pesquisa Agropecu\'{a}ria, Rua Jardim Bot\^{a}nico 
1024, 22460-000 Rio de Janeiro, Rio de Janeiro, Brazil}

\author{J. Maziero}
\affiliation{Centro de Ci\^{e}ncias Naturais e Humanas, Universidade Federal do ABC, R.  Santa Ad\'{e}lia 166, 09210-170 Santo Andr\'{e}, S\~{a}o Paulo, Brazil}

\author{L. C. C\'{e}leri}
\affiliation{Centro de Ci\^{e}ncias Naturais e Humanas, Universidade Federal do ABC, R.  Santa Ad\'{e}lia 166, 09210-170 Santo Andr\'{e}, S\~{a}o Paulo, Brazil}

\author{D. O. Soares-Pinto}
\affiliation{Instituto de F\'{\i}sica de S\~{a}o Carlos, Universidade de S\~{a}o Paulo, Caixa Postal 369, 13560-970 S\~{a}o Carlos, S\~{a}o Paulo, Brazil}

\author{E. R. deAzevedo}
\affiliation{Instituto de F\'{\i}sica de S\~{a}o Carlos, Universidade de S\~{a}o Paulo, Caixa Postal 369, 13560-970 S\~{a}o Carlos, S\~{a}o Paulo, Brazil}

\author{T. J. Bonagamba}
\affiliation{Instituto de F\'{\i}sica de S\~{a}o Carlos, Universidade de S\~{a}o Paulo, Caixa Postal 369, 13560-970 S\~{a}o Carlos, S\~{a}o Paulo, Brazil}

\author{R. S. Sarthour}
\affiliation{Centro Brasileiro de Pesquisas F\'{\i}sicas, Rua Dr. Xavier Sigaud 150, 22290-180 Rio de Janeiro, Rio de Janeiro, Brazil}

\author{I. S. Oliveira}
\affiliation{Centro Brasileiro de Pesquisas F\'{\i}sicas, Rua Dr. Xavier Sigaud 150, 22290-180 Rio de Janeiro, Rio de Janeiro, Brazil}

\author{R. M. Serra}
\email{serra@ufabc.edu.br}
\affiliation{Centro de Ci\^{e}ncias Naturais e Humanas, Universidade Federal do ABC, R.  Santa Ad\'{e}lia 166, 09210-170 Santo Andr\'{e}, S\~{a}o Paulo, Brazil}

\pacs{03.67.-a, 03.65.Ta, 03.67.Ac, 03.65.Ud}

%--------------------------------------------------------------------------------------------------------------------------------------
\begin{abstract}
The quantification of quantum correlations (other than entanglement) usually entails laboured numerical optimization procedures also demanding quantum state tomographic methods. Thus it is interesting to have a laboratory friendly witness for the nature of correlations. In this Letter we report a direct experimental implementation of such a witness in a room temperature nuclear magnetic resonance system. In our experiment the nature of correlations is revealed by performing only few local magnetization measurements. We also compare the witness results with those for the symmetric quantum discord and we obtained a fairly good agreement.
\end{abstract}

\maketitle
%--------------------------------------------------------------------------------------------------------------------------------------
Nonlocality \cite{Bell} and entanglement \cite{Werner} of composed systems are distinguishing features of the quantum domain. Nevertheless, it is the possibility of locally broadcasting \cite{Piani} the state of a multiparticle system that broadly defines the nature of its correlations. Remarkably, even separable (nonentangled) states can be quantum correlated. This kind of quantumness has an important role that is not only related to fundamental physical aspects but also concerning applications in quantum information processing \cite{Datta,Lanyon,Diogo,Terno} and communication \cite{Piani,Cavalcanti},  thermodynamics \cite{Oppenheim}, quantum phase transitions\cite{Sarandy}, and biological systems \cite{Bradler}. 

There are several unique aspects of quantum physics that discern it from classical theories. 
One of particular relevance to quantum information science is the impossibility of creating
a perfect copy of an unknown quantum state \cite{Nielsen}.
This fact is employed in some quantum cryptographic protocols \cite{Nielsen} and, 
when extended to multipartite mixed states \cite{Piani}, can be used to classify the aspects of correlations 
in a composed system as classical or quantum. 
Let us consider a bipartite system described by the density operator $\rho$ and shared by parts 
$a$ and $b$, with respective Hilbert spaces $\mathcal{H}^{a}$ and $\mathcal{H}^{b}$.
The correlations in state $\rho$ are said to be locally broadcast if there are auxiliary systems 
$\mathcal{H}^{a_{1}},\mathcal{H}^{a_{2}},\mathcal{H}^{b_{1}},\mathcal{H}^{b_{2}}$ 
and local operations (completely positive, trace-preserving linear maps)
$\Lambda^{a}:\mathcal{H}^{a}\rightarrow\mathcal{H}^{a_{1}}\otimes
\mathcal{H}^{a_{2}}$ and $\Lambda^{b}:\mathcal{H}^{b}\rightarrow\mathcal{H}^{b_{1}}\otimes
\mathcal{H}^{b_{2}}$ such that $\rho^{a_{1}a_{2}b_{1}b_{2}}=\Lambda^{a}\otimes\Lambda^{b}(\rho)$ 
with $I(\rho^{a_{1}b_{1}})=I(\rho^{a_{2}b_{2}})$, where the quantum mutual information\textemdash
$I(\rho^{xy})=S(\rho^{x})+S(\rho^{y})-S(\rho^{xy})$, 
$S(\rho)=-\mbox{tr}(\rho\log_{2}\rho)$, and $\rho^{x}=\mbox{tr}_{y}(\rho^{xy})$\textemdash
quantifies all the correlations between the systems $x$ and $y$. It turns out that the correlations in a bipartite system 
can be locally broadcast, and therefore have a classical nature, if and only if the system's state can be written as \cite{Piani}

\begin{equation}
 \rho_{cc}=\sum_{i=1}^{\mbox{dim}\mathcal{H}_{a}}\sum_{j=1}^{\mbox{dim}\mathcal{H}_{b}}p_{i,j}
|\alpha_{i}\rangle\langle \alpha_{i}|\otimes|\beta_{j}\rangle\langle \beta_{j}|,
\label{rhoCC}
\end{equation}
where $\{|\alpha_{i}\rangle\}$ and $\{|\beta_{j}\rangle\}$ are orthonormal basis for the subsystems state spaces 
$\mathcal{H}^{a}$ and $\mathcal{H}^{b}$, respectively, and $\{p_{i,j}\}$ is a probability distribution. 

The class of states in equation (\ref{rhoCC}) is contained in the  
set of separable states \textemdash those states that can be generated via local operations
coordinated by communicating classical bits\textemdash whose more general form is $\rho_{sep}=\sum_{i}p_{i}\rho_{i}^{a}\otimes\rho_{i}^{b}$, 
where $\{p_{i}\}$ is a probability distribution and $\rho_{i}^{a}$($\rho_{i}^{b}$) 
is a valid density operator for the subsystem $a(b)$. 
There are separable states that cannot be cast in terms of orthogonal local basis as 
those given in equation (\ref{rhoCC}) and, therefore, present non-classical correlations. 
Underneath such states lies a non-classicality beyond the entanglement-separability paradigm,
which can be quantified by a departure between classical and quantum versions of information theory. 
One of the most popular measures for this kind of non-classicality is the \textit{quantum discord} \cite{Ollivier}. 
This quantifier has been receiving a great deal of attention \cite{Diogo, Terno, Cavalcanti,Maziero5}. 
And it was proposed as a figure of merit for the quantum 
advantage in some computational models without or with little entanglement \cite{Datta,Lanyon}.

\begin{figure}[htb]
\centering
\includegraphics[scale=0.38]{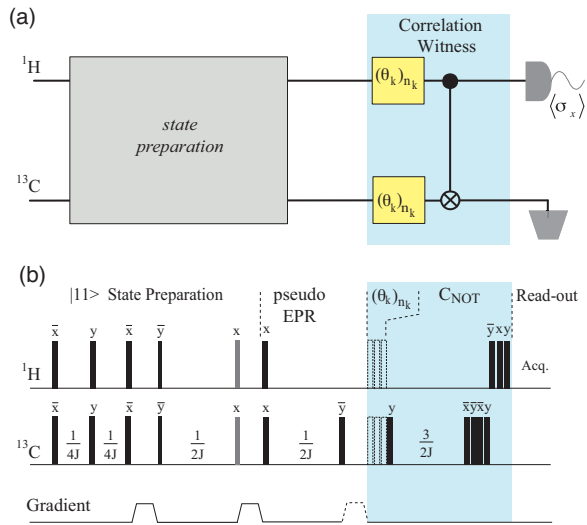}
\caption{(Color online) (a) Schematic representation of the operation sequence used to witness the non-classical nature of correlations. (b) Equivalent pulse sequence employed in our experiment. The thicker filled bars represent $\pi/2$ pulses, the thinner bars indicate $\pi/4$ pulses, and the grey bars indicate $\pi/6$ pulses with the phases as shown (negative pulse phases are described by a bar over the symbol).  The pulses represented as unfilled dashed bars are modified to achieve the different rotations necessary for the witness protocol. The dashed gradient pulse is applied to obtain the classically correlated Bell-diagonal state. The time periods $\frac{1}{2J}$, $\frac{3}{2J}$, and $\frac{1}{4J}$ represent free evolutions under the $J$ coupling \cite{supplementary}.}
\label{fig1} 
\end{figure}

In general, measures of non-classical correlations involve complete knowledge of the 
system's state followed by extremization procedures. In the laboratory, the first task is implemented by quantum 
state tomographic methods and the second one is carried out by additional numerical manipulations. 
These procedures are demanding and propagate the unavoidable experimental errors. 
This observation motivates the search for alternatives regarding the classification of correlations in quantum states.
Once the nature of these correlations somehow determines what can and cannot be done with a given system, 
it is sometimes enough to know whether the correlations in that system have a classical or a quantum nature. 
To accomplish this last task it is convenient to have an observable witness for the quantumness of correlations in the system. 
However, as the state space of classical correlated systems is not convex, a linear witness cannot be used in general, and 
we have to take advantage of a non-linear witness. For a wide class of two-qubit systems,
$\rho=\left(\mathbb{I}^{ab}
+\sum_{i=1}^{3}(\mathcal{A}_{i}\sigma_{i}^{a}\otimes\mathbb{I}^{b}+\mathcal{B}_{i}\mathbb{I}^{a}\otimes \sigma_{i}^{b}
+\mathcal{C}_{i}\sigma_{i}^{a}\otimes \sigma_{i}^{b})\right)/4,$
a sufficient condition for classicality of correlations is \cite{Maziero4}
 \begin{equation}
  W_{\rho}=\sum_{i=1}^{3}\sum_{j=i+1}^{4}\left|\langle O_{i}\rangle_{\rho}\langle O_{j}\rangle_{\rho}\right|=0,
\label{witness}
 \end{equation}
with $O_{i}=\sigma_{i}^{a}\otimes \sigma_{i}^{b}$ for $i=1,2,3$ and 
$O_{4}=\sum_{i=1}^{3}(z_{i}\sigma_{i}^{a}\otimes\mathbb{I}^{b}+w_{i}\mathbb{I}^{a}\otimes \sigma_{i}^{b})$. 
The $\sigma_{i}^{a(b)}$ is the $i$th component of the Pauli operator in subsystem $a (b)$. $\mathcal{A}_{i},\mathcal{B}_{i},z_{i},w_{i}\in \Re$ with 
$z_{i},w_{i}$ randomly chosen and constrained such that $\sum_{i}z_{i}^{2}=\sum_{i}w_{i}^{2}=1$. 
For the so-called Bell-diagonal class of states, 
$\rho_{bd}=\left(\mathbb{I}^{ab}+\sum_{i=1}^{3}\mathcal{C}_{i}\sigma_{i}^{a}\otimes \sigma_{i}^{b}\right)/4$, 
$W_{\rho_{bd}}=0$ is also a necessary condition for the absence of quantumness in the correlations of the composite system (in this case  $\langle O_{4}\rangle_{\rho_{bd}}=0$ \cite{Maziero4}). 
We can easily verify that the observables in equation (\ref{witness}) can be written in terms of one component of the 
magnetization in one subsystem as  $\langle O_{i}\rangle_{\rho}=\langle\sigma_{1}^{a}\otimes\mathbb{I}^{b}\rangle_{\xi_{i}}$, with $\xi_{i}=U_{a\rightarrow b}\left[R_{n_{i}}(\theta_{i})\rho R_{n_{i}}^{\dagger}(\theta_{i})\right]U_{a\rightarrow b}$, 
where $R_{n_{i}}(\theta_{i})=R_{n_{i}}^{a}(\theta_{i})\otimes R_{n_{i}}^{b}(\theta_{i})$, and 
$R_{n_{i}}^{a(b)}(\theta_{i})$ is a local rotation by an angle $\theta_{i}$ around direction $n_{i}$, with 
$\theta_{1}=0$, $\theta_{2}=\theta_{3}=\pi/2$, $n_{2}=y$, and $n_{3}=z$. $U_{a\rightarrow b}$  
is the controlled-NOT gate with the subsystem $a$ as control.

\begin{figure}[htb]
\centering
\includegraphics[scale=0.31]{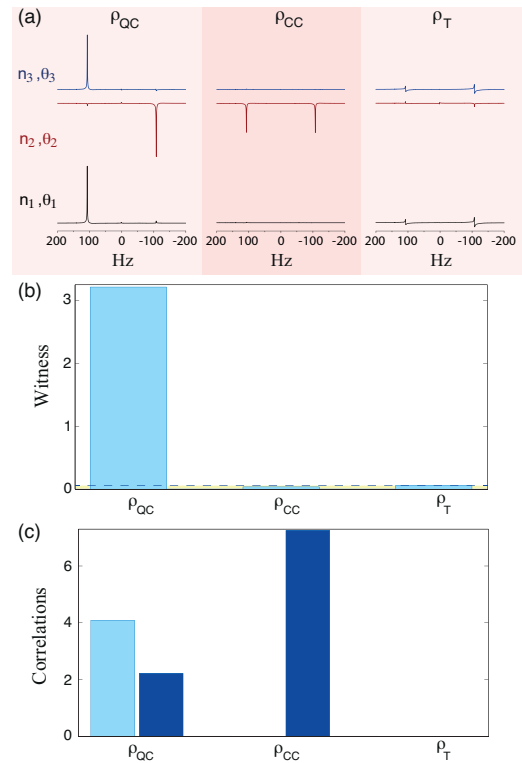}
\caption{(Color online) (a) The $^{1}$H spectra (normalized by the thermal equilibrium state spectrum) obtained after the witness circuit execution (with rotations $R_{n_{1}}$, $R_{n_{2}}$, and $R_{n_{3}}$), (b) witness expectation value, (c) quantum discord (light blue columns) and classical correlation (dark blue columns) measured in three different initial states, $\rho_{QC}$ quantum correlated, $\rho_{CC}$
classically correlated, and $\rho_{T}$ thermal equilibrium state. The dashed line represents the experimental
error bound for determination of classically correlated (zero discord) states. The witness was measured directly performing the circuit depicted in Fig 1 (a), while the classical correlation and the symmetric quantum discord was computed after full QST and numerical extremization procedures. The correlations are displayed in units of ($ \varepsilon^{2} / \ln2$)bit \cite{supplementary}.}
\label{fig2} 
\end{figure}

\begin{figure}[htb]
\centering
\includegraphics[scale=0.30]{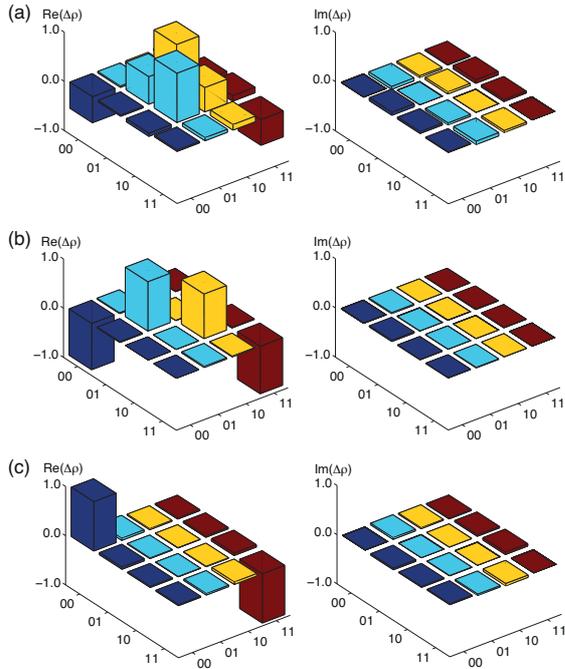}
\caption{(Color online) Real (left) and imaginary (right) parts of the deviation matrix elements reconstructed by QST for the two initial prepared Bell-diagonal states: (a) $\rho_{QC}$ quantum correlated (equivalent to  $\mathcal{C}_{1}=2\varepsilon$, $\mathcal{C}_{2}=2\varepsilon$, and $\mathcal{C}_{3}=-2\varepsilon$); (b) $\rho_{CC}$ classically correlated (equivalent to  $\mathcal{C}_{1}=0$, $\mathcal{C}_{2}=0$, and $\mathcal{C}_{3}=-4\varepsilon$); and also for (c) $\rho_{T}$ the thermal equilibrium state. The deviation matrix elements are displayed in the usual computational basis, where $\left| 0 \right\rangle$ and $\left| 1 \right\rangle$ represent the eigenstates of $\sigma_{z}$ for each qubit. The accuracy of prepared initial states can be estimated by the normalized trace distance from the ideal ones, $\delta(\rho_{ideal},\rho_{prep.})/\varepsilon = \mbox{tr}\left| \Delta\rho_{ideal} - \Delta\rho_{prep.}\right|/2 \approx 0.1$ (for both $\rho_{QC}$ and $\rho_{CC}$).}
\label{fig3} 
\end{figure}

We experimentally implemented the aforementioned witness using the room temperature nuclear magnetic resonance (NMR) system. In this scenario the qubits (quantum bits) are encoded in nuclear spins and they are manipulated by radio-frequency (rf) pulses. Unitary operations are achieved by suitable choice of pulse amplitudes, phases and durations, and the transverse magnetizations are obtained  directly from the NMR signal \cite{NMR}. The state of the two-qubit system is described by a density matrix in the high temperature expansion (where entanglement was ruled out), which takes the form $\rho=\mathbb{I}^{ab}/4+\varepsilon\Delta\rho$, with $\varepsilon=\hbar\omega_{L}/4k_{B}T\sim10^{-5}$ as the ratio between the magnetic and thermal energies and $\Delta\rho$ as the deviation matrix \cite{Nielsen,NMR}. A carbon-13 enriched chloroform (CHCl$_{3}$) solution at $25^{\circ}$C was used in the experiments, with the two qubits being encoded in the $^{1}$H and $^{13}$C spin-$1/2$ nuclei. In order to experimentally demonstrate the witnessing protocol, two initial states were prepared by mapping them into the deviation  matrix using the general pulse sequence scheme as shown in Fig. \ref{fig1}. The first state corresponds to a quantum correlated Bell-diagonal state, which is obtained from the thermal equilibrium by applying the pulse sequence for producing the pseudo-pure state $\left|11\right\rangle$, followed by the pulses that implement a pseudo-EPR gate \cite{chuang1998PRSLA}, see Fig. \ref{fig1}  \cite{supplementary}. The second state is a classically correlated Bell-diagonal state, obtained by applying a z-gradient pulse after the aforementioned pulse sequence. The witnessing of the thermal equilibrium state was also performed as a reference. The experimental procedure depicted in Fig. \ref{fig1} was ran three times for each initial state in order to measure the magnetization $\langle\sigma_{1}^{a}\rangle_{\xi_{i}}$ in the states $\xi_{i}$ that leads to the two-point correlation functions $\langle\sigma_{i}^{a}\otimes\sigma_{i}^{b}\rangle _{\rho}$.  So, the witness given in equation (\ref{witness}) is directly measured (Fig. \ref{fig2}).

\begin{figure}[htb]
\centering
\includegraphics[scale=0.35]{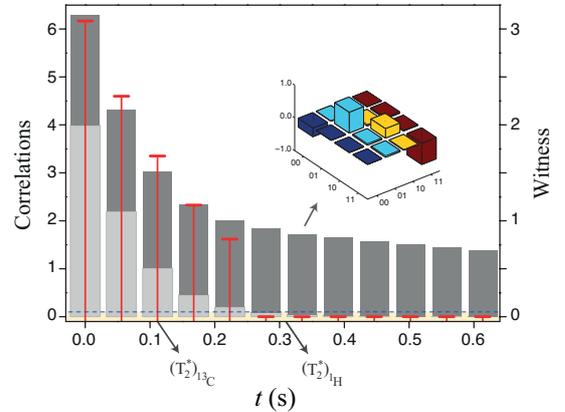}
\caption{(Color online) Witness and correlations decoherence dynamics. The panel displays the measured witness and computed correlations  for  $\rho_{QC}$ [Fig. 3 (a)] relaxed during a time interval, $t_{n}=n \delta t$ ($ \delta t = 55.7$ ms, $n=0,2,...,11$), before performing the witness measurement protocol. The red tick bars represent the witness expectation value (scale on the right), the dark grey section represents the amount of classical correlation, the light grey section represents the symmetric quantum discord, the entire grey bars (light and dark sections summed) display the quantum mutual information (scale on the left).  The classicality bound is represented by the blue dotted line. The inset image shows the real part of the deviation matrix elements reconstructed by QST for an intermediate classically correlated state. The effective transversal relaxation times are $T_{2}^{\ast}=0.31$ s and $T_{2}^{\ast}=0.12$ s, for $^{1}$H and $^{13}$C nuclei, respectively. The correlations are displayed in units of ($ \varepsilon^{2} / \ln2$)bit.}
\label{fig4} 
\end{figure}

For the sake of comparison, we performed a full quantum state tomography (QST) \cite{long2001} of the initial states (displayed in Fig. \ref{fig3}) and computed, from these data, the symmetric quantum discord \cite{Diogo,Maziero3}  \textemdash $Q(\rho^{ab})=I(\rho^{ab})-\max_{\left\{\Pi_{i}^{a},\Pi_{j}^{a}\right\}}\mathcal{I}(\chi^{ab})$, where $\mathcal{I}(\chi^{ab})$ is the measurement-induced mutual information \textemdash and its classical counterpart \cite{supplementary} present in each state. The results are shown in Fig. \ref{fig2}. The correlation quantifiers are computed from the experimentally reconstructed deviation matrix in the leading order in $\varepsilon$, following the approach introduced in Refs. \cite{Diogo,supplementary}. Since the error in the witness expectation value depends on many parameters (i.e., signal-to-noise ratio and residual rf pulse sequence imperfections), we used as a reference the thermal equilibrium state, which is supposed to have no correlations at room temperature \cite{Diogo}. The witness measured for this state ($W_{\rho_{T}}$) was about $0.05$, which is assumed to be the error margin for our experiment. This introduces the bound shown in Fig. \ref{fig2} (b)  for a classically correlated (zero discord) state. 

The witness measured for the three initial states is displayed in Fig. \ref{fig2} (b). For the quantum correlated Bell-diagonal state the witness ($W_{\rho_{QC}}$) is found to be about $3.13$ (far above the $0.05$ bound), while for the classical correlated Bell-diagonal state ($W_{\rho_{CC}}$) it is about $0.04$, i.e., within the classicality cut-off limit. In fact, the witness works perfectly in the present setup, in the sense that it easily sorts out quantum and classically correlated states. Figure \ref{fig2} (c) also displays the quantum discord computed from the experimentally reconstructed deviation matrices using the approach introduced in Ref. \cite{Diogo}. As can be seen, the result for the quantum discord is in agreement with the witness, but the former is obtained after full QST and numerical extremization procedures. Finally, we followed the decoherence dynamics of the witness, by letting the state $\rho_{QC}$ evolves freely during a time period $t_{n}$, after this decoherent evolution we performed the witness circuit, and also a QST in order to compare the witness results with those for correlation quantifiers. The noise spin environment causes loss of phase relations among the energy eigenstates and exchange of energy between system and environment, resulting in relaxation to a Gibbs ensemble. In Fig. 4 we observe, that, in the course of the witness and correlations evolution, the non-classicality is diminished until reaching an only classically correlated state. This occurs near the $^{1}$H effective transversal relaxation time. After such evolution period there are just reminiscent classical correlations, which are also diminished resulting in an uncorrelated state (the room temperature thermal equilibrium state) after the spin-lattice relaxation time. Again, we obtain a fairly good agreement between the witness expectation values and the correlation quantifiers.     

Summarizing, we presented a direct experimental implementation of a witness for the quantumness of correlations (other than entanglement)  in a composite system. Our work showed that it is possible to infer the nature of the correlations in a bipartite system performing only few local measurements over one of the subsystems (just three measurements for both $\rho_{QC}$ and $\rho_{CC}$).  The witness presented in Eq. (\ref{witness}) was generalized to higher-dimensional systems \cite{Ma}. Therefore, the methods employed here can also be easily applied for witnessing correlations in systems with dimensions higher than two. Our strategy precludes the demanding tomographic state reconstruction and the numerical extremization methods included in quantum correlation quantifiers (like quantum discord). This method offers a versatile test-bed for the nature of a composite system that can be applied to other experimental physical contexts. Moreover, in such a proof of principle, we showed that non-classical correlations can be present even in highly mixed states as those in room temperature magnetic resonance experiments.

\begin{acknowledgements}
The authors acknowledge financial support from UFABC, CNPq, CAPES, FAPESP, and FAPERJ. This work was performed as part of the Brazilian National Institute of Science and Technology for Quantum Information (INCT-IQ). RMS thanks S. P. Walborn for insightful discussions.
\end{acknowledgements}

\textit{Note added}.\textemdash After the submission of this Letter, a related work has appeared  \cite{Laflamme}, which employs other methods to witness nonclassicality in an NMR system.
%--------------------------------------------------------------------------------------------------------------------------------------

\appendix

\section{\textbf{Supplemental Material - Details on experimental and calculation procedures}}

\subsection{\textbf{NMR experiments}}

Nuclear magnetic resonance (NMR) experiments were performed on a two-qubit system comprised by nuclear spins of $^1$H and $^{13}$C atoms in a carbon-13 enriched chloroform molecule (CHCl$_{3}$). The sample was prepared by mixing 100 mg of 99 \% $^{13}$C-labelled CHCl$_{3}$ in 0.2 ml of 99.8 \%  CDCl$_{3}$ in a  5 mm NMR tube. Both samples were provided by the Cambridge Isotope Laboratories - Inc. NMR experiments were performed at 25$^{\circ}$ C using a Varian 500 MHz Premium Shielded ($^{1}$H frequency), at the Brazilian Centre for Physics Research (CBPF). A Varian 5mm double resonance probe-head equipped with a magnetic field gradient coil was used.

 The rotating frame nuclear spin Hamiltonian accounting for the relevant NMR interactions reads \cite{NMR-2}
 
\begin{widetext}
\begin{eqnarray}
\mathcal{H} &=&-\left( \omega _{H}-\omega _{rf}^{H}\right) \mathbf{I}%
_{z}^{H}-\left( \omega _{C}-\omega _{rf}^{C}\right) \mathbf{I}_{z}^{C}+2\pi
J\mathbf{I}_{z}^{H}\mathbf{I}_{z}^{C} \nonumber \\
&&+\omega _{1}^{H}\left( \mathbf{I}_{x}^{H}\cos \varphi ^{H}+\mathbf{I}_{y}^{H}\sin \varphi
^{H}\right) +\omega _{1}^{C}\left( \mathbf{I}_{x}^{C}\cos \varphi ^{C}+\mathbf{I}%
_{y}^{C}\sin \varphi ^{C}\right),  \label{Hamiltoniano}
\end{eqnarray}
\end{widetext}
where $\mathbf{I}_{\alpha}^{H}\left(\mathbf{I}_{\beta}^{C}\right)$ is the spin
angular momentum operator in the $\alpha,\beta=x,y,z$ direction for $^{1}$H ($^{13}$C);   $\varphi ^{H}\left(\varphi ^{C}\right)$ defines the direction of the rf field and $\omega _{1}^{H}\left(\omega _{1}^{C}\right)$ is the intensity of  RF pulse for $^{1}$H  ($^{13}$C)  nuclei.

The first two terms describe the free precession of  $^{1}$H and $^{13}$C nuclei about the external constant magnetic field B$_{0}$ with frequencies $\omega _{H}/2\pi \approx $ 500 MHz and $\omega _{C}/2\pi \approx $ 125 MHz, respectively.  The third term describes a scalar  coupling between $^{1}$H  and $^{13}$C nuclei at  $J \approx 215.1$ Hz. The fourth and fifth terms represent the radio frequency (rf) field that may be applied to  $^{1}$H  and $^{13}$C , respectively. The $\pi$/2 pulse has a time setup of 7.4 $\mu$s for $^{1}$H and 9.6 $\mu$s for $^{13}$C at transmitter channel; or a time setup of 7.9 $\mu$s for $^{1}$H and 10.2 $\mu$s  for $^{13}$C at decoupler channel.

Spin-lattice relaxation times (T$_{1}$) of  2.5 s and 7 s for $^{1}$H and $^{13}$C, respectively, were measured using the inversion recovery pulse sequence. The recycle delay was set to $40$ s in all experiments. The effective transversal relaxation times, measured as the inverse of the spectrum line width at halt maximum, were found to be $T_{2}^{\ast}=0.31$ s and $T_{2}^{\ast}=0.12$ s for $^{1}$H and $^{13}$C nuclei, respectively.

In order to experimentally demonstrate the witnessing protocol, two initial states were prepared by mapping them into the deviation density matrix using the general pulse sequence scheme shown in Fig. 1 of the main text. The first state corresponds to a quantum correlated Bell-diagonal state, which is obtained from the thermal equilibrium by applying the pulse sequence for producing the pseudo-pure state $\left|11\right\rangle$, followed by the pulses that implement a pseudo EPR gate \cite{chuang1998PRSLA-2}. It is worth mentioning that, despite a true EPR gate would produce the same state, there is no need of using it in our propose, since the same kind of deviation density matrix can be produced with a pseudo EPR gate, with the advantage of using a smaller number of pulses. A classically correlated Bell-diagonal state, which is obtained by applying a z-gradient pulse after the aforementioned pulse sequence, was also prepared. Finally, the witnessing of the thermal equilibrium state was performed as a reference for error estimation of the whole procedure.  The y rotation in the witness protocol was implemented directly by a single rf pulse, while for the z rotation the pulse sequence $\left(\frac{\pi}{2}\right)_{-y}$-$\left(\frac{\pi}{2}\right)_{x}$-$\left(\frac{\pi}{2}\right)_{y}$ was applied in both nuclei. The controlled-NOT gate was achieved by the pulse sequence $\left(\frac{\pi}{2}\right)_{y}^{C}$-$U\left(\frac{3}{2J}\right)$-$\left(\frac{\pi}{2}\right)_{-x}^{C}$-$\left(\frac{\pi}{2}\right)_{-y}^{C}$-$\left(\frac{\pi}{2}\right)_{-x}^{C}$-$\left(\frac{\pi}{2}\right)_{y}^{C}$-$\left(\frac{\pi}{2}\right)_{-y}^{H}$-$\left(\frac{\pi}{2}\right)_{x}^{H}$-$\left(\frac{\pi}{2}\right)_{y}^{H}$, where the super indices states for nucleus where the pulse is applied and $U\left(\frac{3}{2J}\right)$ represents a free evolution under $J$ coupling. In this case, a pulse sequence that correctly implement a true CNOT gate, i.e., with the correct phases, are indeed necessary to produce the correct output for the witness.  The protocol described in Refs. \cite{long2001-2,teles-2} was used for quantum state tomography (QST). As described above, the experimental procedure depicted in Fig. 1 of the main text is ran three times for each initial state in order to measure the magnetization  $\langle\sigma_{1}^{H}\rangle_{\xi_{i}}$ in the states $\xi_{i}$ that leads to the two-point correlation functions $\langle\sigma_{i}^{H}\otimes\sigma_{i}^{C}\rangle _{\rho}$.  So, the witness given in equation (2) of the main text is directly measured.  

\subsection{\textbf{Measures of correlations}} 

The correlations in the high-mixed state find in the NMR context, 
$\rho^{ab}=\mathbb{I}^{ab}/4+\varepsilon\Delta\rho$ ($\varepsilon\approx 10^{-5}$), 
are computed by expanding the symmetric version of quantum discord \cite{Symmetric-2,Diogo-2} 
\begin{equation}
Q(\rho^{ab})=I(\rho^{ab})-\max_{\left\{\Pi_{i}^{a},\Pi_{j}^{a}\right\}}\mathcal{I}(\chi^{ab}), 
\label{QC}
\end{equation}
where the quantum mutual information (expanded in the leading order in $\varepsilon$), is given by 
\[
I(\rho^{ab})\approx\frac{\varepsilon^{2}}{\ln2}\left\{2\mbox{tr}[(\Delta\rho^{ab})^{2}]-\mbox{tr}\left[(\Delta\rho^{a})^{2}\right]-\mbox{tr}\left[(\Delta\rho^{b})^{2}\right]\right\},
\]
and the measurement-induced mutual information is 
\[
\mathcal{I}(\chi^{ab})\approx\frac{\varepsilon^{2}}{\ln2}\left\{2\mbox{tr}\left[(\Delta\chi^{ab})^{2}\right]-\mbox{tr}\left[(\Delta\chi^{a})^{2}\right]-\mbox{tr}\left[(\Delta\chi^{b})^{2}\right]\right\},  
\]
with $\chi^{ab}=\mathbb{I}^{ab}/4+\varepsilon\Delta\chi^{ab}$ as the state obtained from $\rho^{ab}$ through a complete projective measurement map ($\Delta\chi^{ab}=\sum_{i,j}\Pi_{i}^{a}\otimes\Pi_{j}^{b}(\Delta\rho^{ab})\Pi_{i}^{a}\otimes\Pi_{j}^{b}$). $\Delta\rho^{a(b)}=\mbox{tr}_{b(a)}\{\Delta\rho^{ab}\}$ is the reduced deviation matrix while $\Delta\chi^{a(b)}$ stands for the reduced measured deviation matrix in the subspace $a(b)$. The classical counterpart of Eq. \ref{QC} is $C(\rho^{AB})=\max_{\left\{\Pi_{i}^{a},\Pi_{j}^{b}\right\}}\mathcal{I}(\chi^{AB})$. It is worth mentioning that $Q(\rho^{ab})=0$ if and only if $\rho^{ab}$ can be cast in terms of local orthogonal basis. In other words, the symmetric quantum discord is zero if and only if $\rho^{ab}$ has only classical correlations or no correlations at all (in this case the correlations present in $\rho^{ab}$ can be locally broadcast \cite{Piani-2}). The aforementioned symmetric correlation quantifiers can be computed directly from the experimentally reconstructed deviation matrix and they  were employed to observe the quantumness of a room temperature NMR quadrupolar system \cite{Diogo-2}.


\begin{thebibliography}{99}

\bibitem{Bell} 
J. S. Bell, \textit{Speakable and Unspeakable in Quantum Mechanics} (Cambridge University Press, Cambridge, 1988).

\bibitem{Werner} 
R. F. Werner, Phys. Rev. A \textbf{40}, 4277 (1989).

\bibitem{Piani} 
M. Piani, P. Horodecki, and R. Horodecki, Phys. Rev. Lett. \textbf{100}, 090502 (2008).

\bibitem{Datta}
A. Datta, A. Shaji,  and C. M. Caves, Phys. Rev. Lett. \textbf{100}, 050502 (2008).

\bibitem{Lanyon} 
B. P. Lanyon \textit{et al.},  Phys. Rev. Lett. \textbf{101}, 200501 (2008).

\bibitem{Diogo}
D. O. Soares-Pinto \textit{et al.}, Phys. Rev. A \textbf{81}, 062118 (2010).

\bibitem{Terno}
A. Brodutch and D. R. Terno, Phys. Rev. A \textbf{83}, 010301(R) (2011).

\bibitem{Cavalcanti}
D. Cavalcanti \textit{et al.}, Phys. Rev. A \textbf{83}, 032324 (2011); V. Madhok and A. Datta, Phys. Rev. A \textbf{83}, 032323 (2011). 

\bibitem{Oppenheim}
J. Oppenheim \textit{et al.},  Phys. Rev. Lett. \textbf{89}, 180402 (2002); W. H. Zurek, Phys. Rev. A \textbf{67}, 012320 (2003).

\bibitem{Sarandy}
M. S. Sarandy, Phys. Rev. A \textbf{80}, 022108 (2009); J. Maziero \textit{et al.},  Phys. Rev. A \textbf{82}, 012106 (2010);  T. Werlang \textit{et al.}, Phys. Rev. Lett. \textbf{105}, 095702 (2010); J. Maziero \textit{et al.},  arXiv:1012.5926 (2010).

\bibitem{Bradler}
K. Br\'adler \textit{et al.}, Phys. Rev. A \textbf{82}, 062310 (2010).

\bibitem {Nielsen}
M. A. Nielsen and I. L. Chuang, 
\textit{Quantum Computation and Quantum Information} (Cambridge University Press, Cambridge, 2000).

\bibitem {Ollivier}H. Ollivier and W. H. Zurek, Phys. Rev. Lett. \textbf{88}, 017901 (2001); L. Henderson and V.  Vedral,  J. Phys. A: Math. Gen. \textbf{34}, 6899 (2001).

\bibitem{Maziero5}
J. Maziero \textit{et al.}, Phys. Rev. A \textbf{80}, 044102 (2009);  J. Maziero \textit{et al.}, Phys. Rev. A \textbf{81}, 022116 (2010);  L. Mazzola, J. Piilo, and S.  Maniscalco, Phys. Rev. Lett. \textbf{104}, 200401 (2010); J.-S. Xu \textit{et al.}, Nat. Commun. \textbf{1}, 7 (2010).

\bibitem{Maziero4}
J. Maziero and R. M. Serra, arXiv:1012.3075.

\bibitem{NMR} 
I. S. Oliveira \textit{et al.}, 
\textit{NMR Quantum Information Processing} (Elsevier, Amsterdam, 2007).

\bibitem{chuang1998PRSLA} 
I. L. Chuang \textit{et al.},  Proc. R. Soc. A \textbf{454}, 447 (1998).

\bibitem {supplementary} See Supplemental Material for a detailed description of the experimental and quantum discord calculation procedures.

\bibitem{long2001} 
G. L. Long, H. Y. Yan, and Y. Sun,  J. Opt. B: Quantum Semiclass. Opt. \textbf{3}, 376 (2001); J. Teles \textit{et al.},  J. Chem. Phys.  \textbf{126}, 154506 (2007).

\bibitem{Maziero3}
J. Maziero, L. C. C\'eleri, and R. M. Serra, arXiv:1004.2082.

\bibitem{Ma}
Z.-H. Ma, Z.-H Chen, and J.-L. Chen, arXiv:1104.0299.

\bibitem{Laflamme}
G. Passante \textit{et al.}, arXiv:1105.2262.

\end{thebibliography}

\begin{thebibliography}{9}

\bibitem{NMR-2} 
I. S. Oliveira \textit{et al.},
\textit{NMR Quantum Information Processing} (Elsevier, Amsterdam, 2007).

\bibitem {fortunato2002-2}E. M. Fortunato \textit{et al.}, J. Chem. Phys. \textbf{116}, 7599 (2002).

\bibitem {chuang1998PRSLA-2}
I. L. Chuang\textit{et al.}, Proc. R. Soc. A \textbf{454}, 447 (1998).

\bibitem {long2001-2}
G. L. Long, H. Y. Yan, and Y.  Sun, J. Opt. B: Quantum Semiclass. Opt.  \textbf{3}, 376 (2001).

\bibitem{teles-2} 
J. Teles, \textit{et al.},  J. Chem. Phys. \textbf{126}, 154506 (2007).

\bibitem {Symmetric-2}
J. Maziero, L. C. C\'{e}leri, and R. M. Serra, arXiv:1004.2082.

\bibitem {Diogo-2}D. O. Soares-Pinto \textit{et al.}, Phys. Rev. A \textbf{81}, 062118 (2010).

\bibitem{Piani-2} 
M. Piani, P. Horodecki, and R. Horodecki, Phys. Rev. Lett. \textbf{100}, 090502 (2008).

\end{thebibliography}
\end{document}